\documentstyle{lamuphys}
\makeatletter
\let\chapter\hid@chapter
\makeatother
\begin{document}

\pagenumbering{arabic}
\title{Observing the galaxy environment of QSOs}

\author{Klaus\, J\"ager\inst{1}, Klaus J.\, Fricke\inst{1}, and
Jochen\, Heidt\inst{2}}

\institute{Universit\"ats-Sternwarte, Geismarlandstra{\ss}e 11,
D-37083 G\"ottingen, Germany
\and
Landessternwarte Heidelberg, K\"onigstuhl, 69117 Heidelberg, Germany} 
%Laboratoire d'Analyse Num\'{e}rique, B\^{a}timent 425,\\
%F-91405 Orsay Cedex, France}

\maketitle

\begin{abstract}
We outline our recently started program to investigate the galaxy environment 
of QSOs, in particular of radio--quiet objects at intermediate redshifts.

\end{abstract}
\section{Studies of QSO environments}
Galaxy environment studies of QSOs and of their host galaxies
are important since tidal interaction with neighbouring galaxies is
thought to be one fundamental mechanism for triggering AGN activity 
(e.g.~Osterbrock(1993)). 
The search for galaxy density enhancements (clusters) 
and/or for the presence of close companions near QSOs in support of this 
scenario is only part of the story. 
In addition, one wants to look for qualitative 
differences in the environments associated with intrinsic QSO
properties (e.g.~host galaxy types, radio--loud/quiet), and
for the evolution with redshift. E.g.~previous studies 
(cf.~Ellingson and Yee (1994), Yee and Ellingson (1993))
for $z<0.6$ show basically that radio--loud QSOs tend
to lie in galaxy clusters while radio--quiet QSOs are found in poorer
environments. On the contrary, a few recent observations 
at $z>1$ (Hutchings et al.~(1995), Hutchings (1995)) show hints  for both
radio--loud and radio--quiet QSOs to reside in compact groups of
galaxies. This seems to imply an evolution in density and composition of the 
cluster environment of radio--quiet QSOs at the intermediate red\-shifts. 
A careful comparison and interpretation of
several studies of this kind seems necessary as principle difficulties exist 
for the 
observation and analysis of QSO environment data, e.g.~the interpretation of 
different statistical methods for detecting and measuring galaxy clustering.

\section{A research at intermediate redshifts}
There are nearly no data of QSO environments within 
$0.6<z<1$. We have started a program on the observation of fields around
radio--quiet QSOs at these redshifts.
This research is part of an extensive QSO environment study over a large 
redshift range.  
Our main goals are the detection of galaxy groups or clusters
around the QSOs, an investigation into the density evolution of the
environments, into the role of the host galaxy types
and of gravitational interactions with companions for the development of QSO
activity. 
The first step of our program is a deep imaging survey 
(multicolor broad/narrow--band) of fields around a large sample of QSOs to 
detect host galaxies, close companion-- and cluster candidates. 
Beforehand all fields were checked
for contamination by known foreground galaxy clusters, radio sources,
bright foreground galaxies or stars. 
As a second step spectroscopy of selected 
targets is intended to verify the physical association of environment 
galaxies with the QSOs and to assess their state of activity. 
A preliminary analysis of our obtained data holds already some promising 
results. A comparison of R-- and I-- band observations of fields around  
the QSOs reveals evidence for the physical association 
of companion galaxies with some QSOs. While the galaxies are easily detected in 
the I--band the comparable R--band image shows these objects to be more 
diffuse and fainter. This is to be expected at redshifts $z>0.75$ due to the
restframe wavelength ranges covered by the selected broad--band filters
(eg.~in the R--band below the 4000 \AA~ break).
Fig.~1 shows e.g.~the compact group of objects near the QSO
2249--0154 at $z=0.83$ (I--band image, 50 min., Calar Alto 2.2m 
telescope). 
11 objects within 20 arcsec or within a 
projected distance of about 300 kpc around the QSO 
can be seen (one being a likely foreground star).
There are some other cases where we found 
close--companion galaxies or even cluster candidates revealing the typical 
shape and size of a cluster to be expected at this redshift.
In summary, the first inspection of our data supports the  
presumption that radio--quiet QSOs reside in a variety of environments at 
intermediate redshifts.
\vspace*{-0.5cm}
\input{psfig}
\begin{figure}
\centerline{\psfig{figure=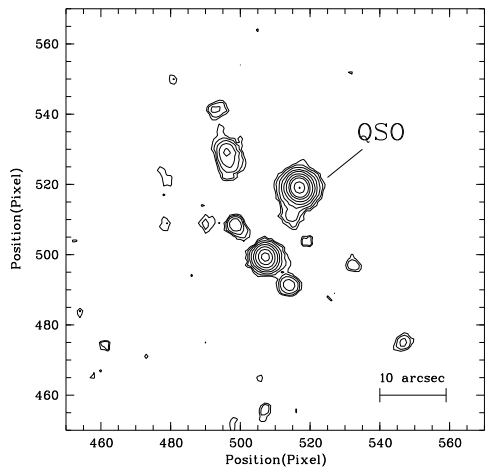,height=6.64cm,width=7cm}}
\vspace{-0.3cm}
\caption{The close environment of the radio--quiet QSO 2249--0154.(cf.~text)}
\end{figure}
\vspace*{-5mm}
%
%
% ---- Bibliography ----
%
%\begin{thebibliography}
\flushleft
{\bf{References:}}\\[2mm]
%
%\bibitem{}{ell:yee}{Ellingson and Yee (1994)}
Ellingson,\, E.,\, Yee,\,H.K.C. (1993):
ApJ Suppl.~{\bf 92}, 33 \\
%
%\bibitem{}{huta}{Hutchings et al. (1995)}
Hutchings,\, J.B., Crampton,\, D., Johnson,\, A. (1995):
AJ {\bf 109}, 73 \\
%
%\bibitem{}{hutb}{Hutchings (1995)}
Hutchings,\, J.B., (1995):
AJ {\bf 109}, 928 \\
%
%\bibitem{}{ost}{Osterbrock (1993)}
Osterbrock,\, D.E. (1990): ApJ 404, 551 \\
%
%\bibitem{}{yee:ell}{Yee and Ellingson (1993)}
Yee,\,H.K.C.,\, Ellingson,\, E. (1993):
ApJ {\bf 411}, 43 \\[2mm]
This work has been supported by the Deutsche Forschungsgemeinschaft (FR 
325/44--1/42--1)
%
%\end{thebibliography}
\end{document}